\documentstyle[twocolumn,epsfig,aps]{revtex}
\begin{document}
\setlength{\topmargin}{-0.05in} \draft
\title{Quantum Zeno subspace and
entangled Bose-Einstein condensates}
\author{M. Zhang$^{1,\dag}$ and L. You$^{1,2}$}
\address{$^1$School of Physics, Georgia Institute of Technology,
Atlanta, Georgia 30332, USA}
\address{$^2$Interdisciplinary Center of Theoretical Studies
and Institute of Theoretical Physics, CAS, Beijing 100080, China}
\date{\today}
\maketitle
\begin{abstract}
We discuss a proposal for the efficient generation of the
maximally entangled atomic N-GHZ state in a spinor-1 condensate by
driving internal state atomic Raman transitions using (classical)
laser fields. We illustrate the dynamics in terms of a quantum
Zeno subspace, and identify the resultant atomic elastic
collision in facilitating the deterministic entanglement creation.
Our proposal can be readily implemented in several laboratories
where ferromagnetic spinor condensates (of $^{87}$Rb atoms) are
investigated.
\end{abstract}

\pacs{03.65.Xp, 03.67.Mn, 03.75.Gg, 03.75.Mn}


The continued success of the research on quantum degenerate atomic gases
has led to intense exploration of their potential application
for quantum information science. A substantial topic of recent
interest is the creation of spin squeezed atomic states that display
multi-particle inseparable correlations \cite{sorensen01,youold}.
While it is debatable how such inseparable {\it quantum
correlations} among identical particles are related to {\it
entanglement} between distinct parties \cite{youi,mil}, it has
nevertheless been fruitful to investigate these unexplored
territories. Maximally entangled states enjoy special attention;
in addition to being states
of maximum inseparable {\it correlations} or {\it entanglement}, they
compose of a special class with which theoretical discussions
can become most transparent.

Recently, a deterministic protocol was suggested \cite{you0209}
for creating maximally entangled pairs,
triplets, quartiles, and other clusters of Bose condensed atoms
starting from a condensate in the Mott insulator state
\cite{greiner02}. The work of \cite{you0209} concerns a single
optical well and involved a small number of atoms. In this limit,
the evolution dynamics of the system is
analytically solvable, which allows the
appropriate timing of the external laser fields
(in comparison with atom elastic collision strength)
to be identified for creating maximal atomic entanglement.

This article clarifies the operating mechanism of the
seemingly simple protocol \cite{you0209}.
More importantly, we investigate its application to
large numbers of condensed atoms. To our surprise, the few atom protocol
with classical laser fields driving atomic Raman transitions
remains efficient in generating massive atomic entanglement
of condensed atoms. The answer, it turns out, can be obtained
rather conveniently understood in terms of a quantum Zeno subspace
\cite{facchi02}. Within this subspace, the effective atom interaction
and the Raman dynamics reduces to a well-known form studied earlier
\cite{molmer}. What is crucial for creating maximally entangled atomic
states is to start from a condensate with a fixed
number of atoms, (be it even or odd \cite{molmer}).

This paper is organized as follows. First, we briefly introduce
our system and the setup for generating maximal entangled atomic
states. Then we provide a step by step
illustration of the underlying mechanism for the creation of such
massively correlated atomic states. We briefly review
the concept of a quantum Zeno subspace \cite{facchi02}. This is
followed by a review of a spin-1 condensate, paying special
attention to the case of ferromagnetic atom-atom interactions
(as for $^{87}$Rb atoms). We then discuss an interesting
decomposition of the interaction Hamiltonian into three sub-SU(2)
spaces as first suggested in \cite{ozgur}. Finally, we show that
the external laser fields in the appropriate Raman configuration
create a quantum Zeno subspace, within which
the effective interaction is of the desired form for generating
spin squeezing and maximal atomic entanglement. We conclude
with some illustrating numerical results.

Our main result can be summarized in Fig. \ref{fig1} below.
We consider a spin-1 ($F=1$ atoms with three Zeeman sub-levels)
condensate with a fixed number (N) of atoms that are all in the
$|+\rangle\equiv |M_F=1\rangle$ or $|-\rangle\equiv
|M_F=-1\rangle$ state initially. We claim that if a strong
off-resonant Raman transition is established between the
$|+\rangle$ and $|-\rangle$ state, then at predictable instants of
the time evolution, the condensate becomes maximally entangled,
i.e. its atomic internal state becomes an N-atom GHZ state
$\propto |+\rangle^{\otimes N}+ |-\rangle^{\otimes N}$.

\begin{figure}
\includegraphics[width=3.in]{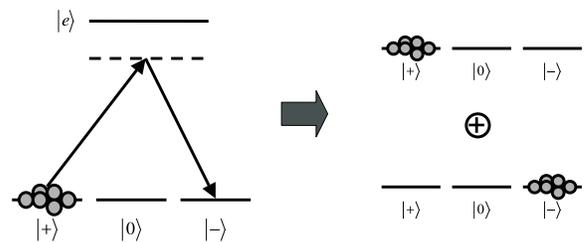}\\
\caption{Illustration of our basic result.} \label{fig1}
\end{figure}

{\bf Quantum Zeno subspace} The quantum measurement process itself can
strongly affect the dynamics of a system. An often cited
example in this regard is the famous quantum Zeno effect
\cite{cook,itano}: {\it a watched pot does not boil}, or in a
physical system, frequent measurements of the initial state hinders
its dynamics to evolve to other states. Recently, Facchi and
Pascazio provided new insights into the Zeno effect, by
reformulating it in terms of an adiabatic theorem within different
sectors of the so-called quantum Zeno subspaces \cite{facchi02},
which are closely related to the powerful decoherence free
subspace in quantum computing studies \cite{knight,lidar}.
In terms of the notation of \cite{facchi02}, the concept of a
quantum Zeno subspace can be illustrated by a system Hamiltonian
\begin{eqnarray}
H_\Omega=H+\Omega H_{\rm meas}. \label{hz}
\end{eqnarray}
In the limit of an infinitely large parameter $\Omega$, the time
evolution of $H_\Omega$ is diagonal with respect to spaces of
$\Omega H_{\rm meas}$, i.e. for any Hamiltonian of the form
(\ref{hz}), the largest term $\Omega H_{\rm meas}$ decomposes the
complete Hilbert space into different subspaces or quantum Zeno
subspaces. The dynamics associated with the small term $H$ is
slaved to slow adiabatic evolutions within each of these
orthogonal subspace.

{\bf Spinor condensate} For ferromagnetic
interactions (e.g. $^{87}$Rb),
as was shown before \cite{sma}, the spatial mode of
different condensate component $\phi_j(\vec r)$ (normalized to
unity) is identically the same $\phi(\vec r)$. Thus, a spin-1
condensate can be described with $a_{j=+,0,-}$
($a_{j=+,0,-}^\dag$), the single atom annihilation (creation)
operator for each spin component state. Apart from constant terms
that depend on the total number of atoms $N=N_++N_0+N_-$,
($N_j=a_j^\dag a_j$), the atom-atom interaction becomes
\cite{pu,ho2,ueda2}
\begin{eqnarray}
H=uL^2,
\label{ham}
\end{eqnarray}
where the coefficient is $u=c_2\int |\phi(\vec
r)|^4d\vec r/2$ ($u<0$ for ferromagnetic interactions).
$c_2={4\pi\hbar^2}(a_2-a_0)/3M$ with $a_2$ and
$a_0$ the respective scattering lengths in the total
spin $F=2$ and $0$ channels of two colliding spin-1 atoms. The pseudo
angular momentum operator $\vec L$ is defined according to the Schwinger
representation, and with its components given by
\begin{eqnarray}
L_+ &&=L_x+iL_y=\sqrt{2}(a_+^{\dagger}a_0+a_0^{\dagger}a_-),\nonumber\\
L_- &&=L_+^{\dagger},\hskip 12pt L_z =a_+^{\dagger}a_+-a_-^{\dagger}a_-.
\end{eqnarray}
It is well known \cite{ozgur} that such an effective
interaction (\ref{ham}) is formally SU(2) symmetric, thus would
have not generated any nonlinear dynamics if the $L_j$'s were the
SU(2) Schwinger representation of two bosonic modes
\cite{kitagawa}. For the spin-1 condensate being
considered here, as was emphasized before \cite{ozgur}, the pseudo
angular momentum operator $\vec L$ does NOT satisfy the Casimir relation
[$L^2\neq N(N+1)$], and therefore can not really be considered
an angular momentum operator \cite{ozgur}. Using the Gell-Mann
decomposition of SU(3) into three SU(2) subspaces $U$, $V$, and
$T$, it was found \cite{ozgur} that
\begin{eqnarray}
L^2=&&4T_z^2+2(V_+U_++V_-U_-)\nonumber\\
&&+(N-\epsilon_+)(N-\epsilon_-)/2-2(Y-Y_0)^2, \label{l2}
\end{eqnarray}
with $\epsilon_\pm=-3/2\pm\sqrt{2}$ and $Y_0=-{N}/{6}-{1}/{4}$. It is
important to note that operators belonging to different SU(2)
subspaces do not
always commute. We adopt the convention that
\begin{eqnarray}
T_+&=&a_+^{\dagger}a_-,\quad T_z=\frac{1}{2}(a_+^\dag a_+-a_-^\dag
a_-),\\
\left(\begin{array}{c} V_+\\U_+
\end{array}\right)&=&a_\pm^{\dagger}a_0, \quad
\left(\begin{array}{c} V_z\\U_z
\end{array}\right)=\frac{1}{2}(a_\pm^\dag a_\pm-a_0^\dag a_0),
\end{eqnarray}
and the hypercharge $Y\equiv (N_++N_--2N_0)/3$, which physically
corresponds to the quadrupole moment of atomic populations
$N_{j=+,0,-}$.

{\bf Effective interaction and maximal entanglement generation}
A Hamiltonian of the form $T_z^2$ [the first term in Eq.
(\ref{l2})] can be used to generate maximally entangled
or spin squeezed states
\cite{sorensen01,youold,molmer,zoller}. To selectively suppress
other terms in the Hamiltonian Eq. (\ref{ham}), we resort to the
idea of quantum Zeno subspace as reviewed earlier. In this case,
to limit the dynamics to the manifold composed of internal states
$|+\rangle$ and $|-\rangle$, we simply Raman couple them with
strong external laser fields, in the form
$\Omega(T_+-T_-)/{2i}=\Omega T_y$ as shown in Fig. \ref{fig1}. In
the spirit of the adiabatic theorem within the Zeno
subspace \cite{facchi02}, our system dynamics is now governed by
\begin{eqnarray}
H'&=&4uT_z^2+\Omega T_y, \label{hp}
\end{eqnarray}
an effective Hamiltonian that is known to generate maximally
entangled states \cite{you0209,molmer}.

The shortest time for creating a maximally entangled state when
initially all atoms are in either $|+\rangle$ or $|-\rangle$ is,
$\pi/4|u|$, a factor of two longer than that from a bare
interaction of the form $H'=4uT_x^2$ \cite{you0209}. This point
becomes clear when we transfer to the interaction picture
$|\psi(t)\rangle_I=\exp(i\Omega T_y t)|\psi(t)\rangle$ as
now $\Omega T_y$ becomes the largest term in the system
Hamiltonian. Equation (\ref{hp}) arises when the limit
$\Omega\gg N|u|$ is satisfied. The effective Hamiltonian in the
interaction picture then becomes
\begin{eqnarray}
H_{\rm int}'&&=e^{i\Omega T_y t}(4uT_z^2)e^{-i\Omega T_y
t}\nonumber\\
&&=4u(T_z\cos\Omega t-T_x\sin\Omega t)^2\nonumber\\
&&\approx 2u(T_z^2+T_x^2) = 2uT^2-2uT_y^2, \label{hint}
\end{eqnarray}
where the approximation is due to averaging over the rapid Rabi
oscillations. The first term $\propto T^2$ in Eq. (\ref{hint}) is
SU(2) symmetric in the two dimensional subspace of $|+\rangle$ and
$|-\rangle$, thus it can be neglected as it only introduces an
overall phase factor. The second term, $\propto T_y^2$
(with a coefficient a factor of 2 smaller), is similar
to the $T_x^2$ term \cite{you0209,molmer}, thus explaining the
dynamic creation of maximally entangled states.
The entanglement remains for the Schr\"{o}dinger
picture based atomic states as the two pictures coincide
at integer periods of the Rabi oscillation.
The rapid time average results in a factor of two reduction of
the effective interaction strength.

We now present selective results from numerical simulations based
on the complete Hamiltonian (\ref{ham}) plus the Raman term
$\Omega T_y$ and compare with the approximate results based on the
adiabatic Hamiltonian (\ref{hp}).

\begin{figure}
\includegraphics[width=3.25in]{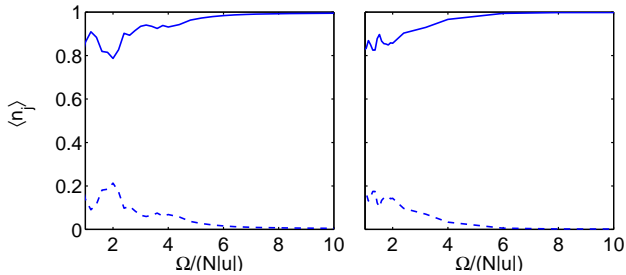}\\
\caption{\label{phase} The average fraction of atoms in the Zeno
subspace $n_z=\langle N_++N_-\rangle/N$ (solid line) and in state
$|0\rangle$, $n_0=\langle N_0\rangle/N$ (dashed line) for $N=10$
(left panel) and 50 (right panel) at $t=\pi/4|u|=5$ (ms).
The qualitative dependence is similar at other times.
$u/(2\pi)=-25$ (Hz) and initially all atoms in state $|+\rangle$.}
\end{figure}

\begin{figure}
\includegraphics[width=3.25in]{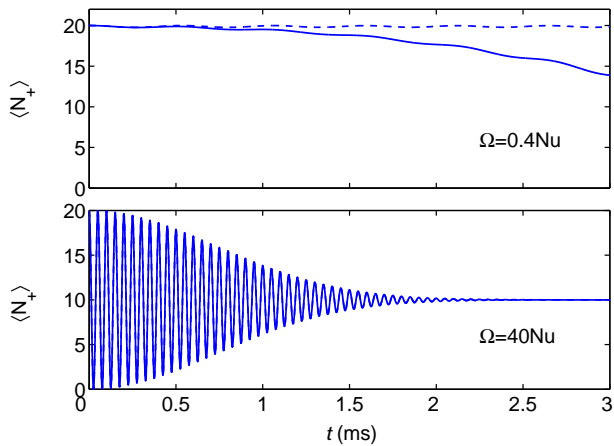}\\%
\caption{\label{aver} The time dependence of $\langle N_+\rangle$
for $N=20$, $u/(2\pi)=-25$ (Hz), and initially all atoms in state $|+\rangle$.
The solid line is from the complete Hamiltonian while the dashed line
is from the adiabatic Hamiltonian in the Zeno subspace.
At $\Omega=40N|u|=40\times (20|u|)=(2\pi)20$ (kHz) (lower panel),
the Zeno subspace is well established.}
\end{figure}

\begin{figure}
\includegraphics[width=3.25in]{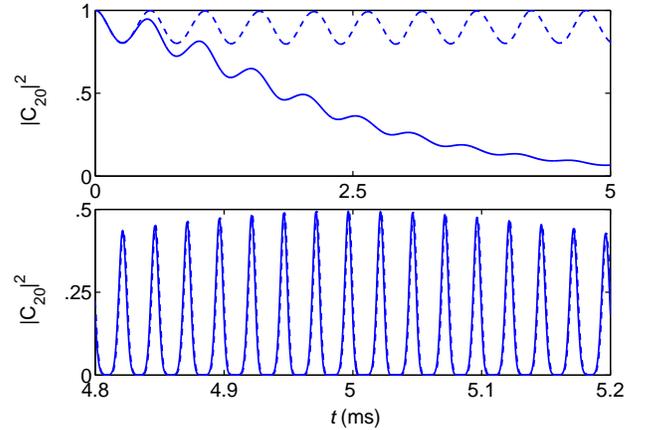}\\%
\caption{\label{prob} The same as in Fig. \ref{aver} but
the probability for all atoms in $|+\rangle$.
$C_{N}=\langle \psi(t)|a_+^{\dag N}|{\rm vac}\rangle/\sqrt{N!}$. Note the
the creation of a N-GHZ state at time $\sim 5$ (ms) (lower panel). }
\end{figure}

Figure \ref{phase} demonstrates the validity of a quantum Zeno
subspace ($|+\rangle$ and $|-\rangle$) with increasing values of
the effective Rabi frequency and for two different atom numbers.
Indeed, when $\Omega$ becomes larger as compared to $N|u|$, the
populations become increasingly localized
within the isospin subspace ($|+\rangle$, $|-\rangle$),
and eventually the Hamiltonian $H'$
correctly describes the dynamics and maximally entangled N-GHZ
states are created at $t=\pi/4|u|$. The creation of a Zeno subspace can
be further illustrated by watching the time dependent population
distributions as in Figs. \ref{aver} and \ref{prob}.

We found that if a strong Raman coupling is
established between states $|+\rangle$ ($|-\rangle$) and
$|0\rangle$, a quantum Zeno subspace composed of atoms in these
two states is also created. However, maximally entangled states do
not occur as the projection of $L^2$ into this space does not
yield the required nonlinear interaction $V_z^2$ ($U_z^2$) \cite{ozgur}.

We now discuss the generation of N-GHZ states when $\Omega\gg
N|u|$. Because the interaction picture Hamiltonian (\ref{hint}) is
similar to that of Molmer's model as in \cite{molmer}, we anticipate similar
dynamical behaviors. We adopt the notion of the maximal entangled
fraction and define the optimized overlap
\begin{eqnarray}
F_{\rm max}&&=\max_{\theta,\phi,\eta}(|_N\langle
E_{\theta,\phi,\eta}|\psi\rangle|^2)\nonumber\\
&&=0.5(|\alpha|+|\beta|)^2,
\end{eqnarray}
of the N-atom wave function $|\psi(t)\rangle$ with an
entangled coherent spin state \cite{cs}
\begin{eqnarray}
|E_{\theta,\phi,\eta}\rangle_N
&&=\frac{1}{\sqrt{2}}(|\theta,\phi\rangle^{\otimes N}
+e^{-i\eta}|\bar{\theta},\bar{\phi}\rangle^{\otimes N}),
\end{eqnarray}
with $\bar{\theta}=\pi-\theta$,
$\bar{\phi}=\pi+\phi$, $\alpha=\langle\theta,\phi|\psi\rangle$,
and $\beta=\langle\bar{\theta},\bar{\phi}|\psi\rangle$. For
$F_{\rm max}$ we note $\eta=\arg(\beta)-\arg(\alpha)$.

\begin{figure}
\includegraphics[width=3.25in]{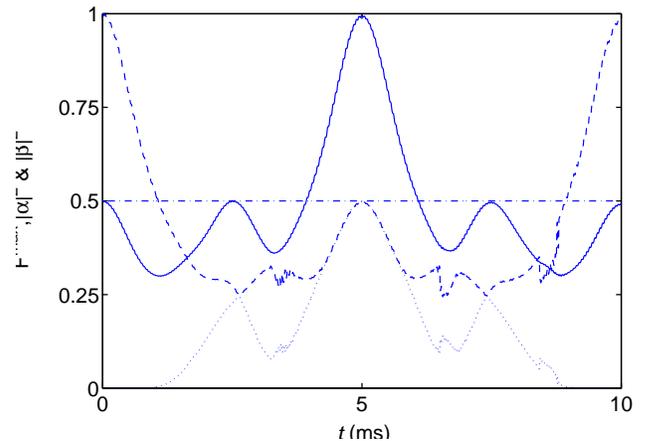}\\
\caption{\label{maximal}The time dependence of $F_{\rm max}$
(solid line), $|\alpha|^2$ (dashed line), and $|\beta|^2$ (dotted
line) for $N=20$, $u/(2\pi)=-25$ (Hz), and $\Omega=(2\pi) 10$ (kHz). }
\label{fig5}
\end{figure}

In Fig. \ref{maximal}, we display the time dependence of $F_{\rm
max}$ with the use of the adiabatic Hamiltonian
(\ref{hp}). We see two interesting regimes. For times up to
$1/(N|u|)$, the wave function is almost a coherent spin state,
so the mean field theory can be used to describe the system dynamics
\cite{zoller}. Around $t=\pi/4|u|=5$ (ms), however, we obtain a high
fidelity entangled coherent spin state, where the mean field
theory completely breaks down. We also find that the
angles $\theta_{\rm max}$, $\phi_{\rm max}$,
and $\eta_{\rm max}$ are essentially constant
in the immediate neighborhood of $t=\pi/4|u|=5$ (ms) (in the interaction
picture), thus a phase sensitive detection of the massive
entanglement can be accomplished without extremely precise
timing \cite{zoller,zhang}.
$F_{\rm max}$ has been proven to be a
useful measure of N-atom entanglement:
a N-atom pure state $|\psi\rangle$ is entangled if $F_{\rm max}>1/2$
(above the dash dotted line in Fig. \ref{fig5}) \cite{bei,npt}.

Finally, we note that the maximally entangled state
$|+\rangle^{\otimes N}+|-\rangle^{\otimes N}$ as generated with
our method is stable against elastic two body collisions, thus is
ideal for studying multiple atom entanglement. Furthermore,
our protocol works, even when there is a non-zero magnetic field
that breaks the level degeneracy for different spin components.
The Zeeman interaction gives rise to
\begin{eqnarray}
H_B=-\hbar\omega_L(a_+^\dag a_+-a_-^\dag a_-)
+\hbar\delta(a_+^\dag a_++a_-^\dag a_-),
\end{eqnarray}
where the linear Zeeman term is proportional to the Larmor
precessing frequency $\omega_L=B\mu_B/\hbar$, with $\mu_B$ the
magnetic dipole moment for state $|\pm\rangle$, and the quadratic
term is proportional to $\delta\propto B^2$, one half the energy
released from the generation of two atoms in state $|0\rangle$
due to the elastic collision of one
atom in state $|+\rangle$ with another one in state
$|-\rangle$. In the quantum Zeno subspace, $a_+^\dag a_++a_-^\dag
a_-=N$, remains a constant, so the quadratic Zeeman term does not
affect the dynamics of generating maximally entangled N-GHZ states.
The linear Zeeman term is $\propto T_z$, which simply induces a
twisted phase on the state $|\pm\rangle$. In the rotating frame, the
atom-atom interaction term remains the same $4uT_z^2$, while the
Raman coupling $\Omega T_y$ now contains time dependent phase
factors $e^{\pm i\omega_Lt}$. Nevertheless the complete system
dynamics remain the same. Therefore we expect our theoretical
protocol to remain effective.

In conclusion, we have proposed a protocol for robustly generating
maximally entangled atomic states in a condensate. We have illuminated
the operating principle in terms of a quantum Zeno
subspace. Our protocol can be implemented in currently available
condensate systems with ferromagnetic interactions, where
the spatial mode functions are identical for each of
the components \cite{sma}. It can also be applied to spin-1
condensates with anti-ferromagnetic interactions
(e.g. $^{23}$Na atoms \cite{mit}), which may possess larger values
of exchange interaction $u$, thus leading to faster generation of
entangled condensates. Although the different equilibrium
mode functions in the latter case can lead to modulational
instability, thus increased decoherence of
the maximally entangled state \cite{sma}. Our protocol starts with
all condensed atoms in either state $|+\rangle$ or $|-\rangle$.
Provided the time scale $1/|u|$ is reasonably short as
compared to the inverse of a typical condensate phonon frequency,
the dynamics for generating entangled condensate is also
mechanically stable, irrespective of whether the
$|+\rangle$ or $|-\rangle$ condensate component is miscible
(for anti-ferromagnetic interactions)
or not (ferromagnetic interactions).

This work is supported by NSF, CNSF, and by a grant from NSA,
ARDA, and DARPA under ARO Contract No. DAAD19-01-1-0667.

\end{document}